\documentclass[aps,prl,twocolumn,showpacs]{revtex4}
\usepackage{amssymb}
\usepackage{amsmath}

\begin{document}

\title{Boson Representation of Spin Operators}

\author{Lasha Tkeshelashvili}
\affiliation {
Andronikashvili Institute of Physics, Tbilisi State University, 
Tamarashvili 6, 0177 Tbilisi, Georgia}

\begin{abstract}
The derivation of the boson representation of spin operators is given which reproduces the Holstein-Primakoff and Dyson-Maleev transformations in the corresponding cases. The suggested formalism allows to address some subtle issues which appear crucial for treating certain class of problems. Moreover, the transformation is suggested which is naturally related to the symmetry of the spin systems. 

\end{abstract}


\maketitle

The boson representation of spin operators was first developed by Holstein and Primakoff in the context of quantum magnetism \cite{holstein1940}. However, due to technical difficulties inherent to the Holstein-Primakoff transformation, Dyson presented an alternative approach to the problem \cite{dyson1956a, dyson1956b}. Based on the Dyson theory, Maleev suggested another representation of spin operators \cite{maleev1958}. In addition, an important extension of the Maleev result was reported in Ref.~\cite{dembinski1964},  where the conjugated Dyson-Maleev transformation is derived. 
Although the Holstein-Primakoff and the Dyson-Maleev transformations represent the basic tools for treating the spin or pseudo-spin dynamics in various fields of quantum theory, there still remain certain subtle points which remain to be addressed. Below, the derivation of the boson representation of spin operators is given which permits to solve such issues, and to find the transformation that is consistent with the group-theoretical treatment of the spin systems \cite{takhtajan2008, miller1972}. 

Let us first recall some relevant relations of the quantum theory of angular momentum. In particular,
the spin operator $\hat{\bf S} = (\hat S^x, \hat S^y, \hat S^z)$ can be introduced through the commutation rules \cite{takhtajan2008}:
\begin{align}
& \hat S^z \hat S^+ - \hat S^+ \hat S^z = \hat S^+ \, , \label{spincommzpl} \\
& \hat S^- \hat S^z - \hat S^z \hat S^- = \hat S^+ \, , \label{spincommminz} \\
& \hat S^+ \hat S^- - \hat S^- \hat S^+ = 2 \hat S^z \, ,\label{spincommplmin}
\end{align}
with $\hat S^{\pm} = \hat S^x \pm i \hat S^y$. Assuming that $\hat S^z$ is diagonal, we have:
\begin{equation}
\hat S^z \vert M,S \rangle = M \vert M,S \rangle \, .
\label{szeigenpr}
\end{equation}
Here $S$ stands for the spin magnitude. The result of operating with $\hat S^+$ and $\hat S^-$ on $\vert M,S \rangle$ follows from the commutation rules. In particular, the straightforward calculations give
\begin{equation}
\hat S^z \hat S^{\pm} \vert M,S \rangle = (M \pm 1)\hat S^{\pm} \vert M,S \rangle \, .
\label{spinupdown}
\end{equation}
That is $\hat S^{\pm} \vert M,S \rangle = \vert M \pm 1,S \rangle$. Nevertheless,  $M = \overline{-S,+S}$ as the spin magnitude is $S$. That implies: 
\begin{align}
& \hat S^+ \vert +S, S \rangle = 0 \, ,  \label{maxspl} \\
& \hat S^- \vert -S, S \rangle = 0 \, .   \label{minsmin}
\end{align}

The expressions for $ \hat{\bf S}^2 = \hat S^{x2} + \hat S^{y2} + \hat S^{z2}$ can be derived from Eqs.~\eqref{spincommzpl}, \eqref{spincommminz} and \eqref{spincommplmin}:
\begin{align}
& \hat{\bf S}^2 = \hat S^+ \hat S^- + \hat S^z (\hat S^z - 1) \, ,  \label{ssqrplmin} \\
& \hat{\bf S}^2 = \hat S^- \hat S^+ + \hat S^z (\hat S^z + 1) \, .   \label{ssqrminpl}
\end{align}
These equations, in combination with Eqs.~\eqref{maxspl} and \eqref{minsmin}, give
\begin{equation}
\hat {\bf S}^2 = S(S+1) \, .
\label{ssqrresult}
\end{equation}

The main idea of the boson representation of spin operators is based on the following assumption \cite{holstein1940, dyson1956a}:
\begin{equation}
\hat S^z = -S +\hat a^+ \hat a \, ,
\label{szboson}
\end{equation}
where the ground state of the spin is chosen to be $ \vert -S, S \rangle$. Moreover, $\hat S^+=\hat S^+(\hat a^+, \hat a)$ and $\hat S^-=\hat S^-(\hat a^+, \hat a)$.
The operators $\hat a$ and $ \hat a^+ $ obey the canonical boson commutation relation \cite{louisell1964}:
\begin{equation}
[\hat a \ ,\hat a^+] = 1 \, .
\label{bosoncomm}
\end{equation}
Note that, the eigenvalues of  $\hat n = \hat a^+ \hat a$ are nonnegative integers $N$. However, since $M = \overline{-S,+S}$, the physical range of the eigenvalues is $0\le N \le2S$.  Therefore, similar to  Eqs.~\eqref{maxspl} and \eqref{minsmin}, we must require that
\begin{align}
& \hat S^+(\hat a^+, \hat a) \vert  2S \rangle = 0 \, ,  \label{maxsplboson} \\
& \hat S^- (\hat a^+, \hat a)\vert 0 \rangle = 0 \, .   \label{minsminboson}
\end{align}
So, the problem reduces to the derivation of $\hat S^+(\hat a^+, \hat a)$ and $\hat S^-(\hat a^+, \hat a)$ that obey the commutation rules \eqref{spincommzpl}, \eqref{spincommminz}, \eqref{spincommplmin} as well as Eqs.~\eqref{maxsplboson} and \eqref{minsminboson}. 

First, let us derive the expression for $\hat S^+(\hat a^+, \hat a)$. This can be done with the help of 
the following general relations \cite{louisell1964}:
\begin{align}
& [\hat a^{\ }, f] = \frac{\partial f}{\ \partial \hat a^+} \, ,  \label{derivpl} \\
& [\hat a^+ , f] = - \frac{\partial f}{\partial \hat a} \, .   \label{deriv}
\end{align}
Here $f \equiv f(\hat a^+,\hat a)$. Taking into account these relations, from Eqs.~\eqref{spincommzpl} and 
\eqref{szboson} we obtain
\begin{equation}
\hat a^+ \frac{\partial \hat S^+}{\partial \hat a^+} - \frac{\partial \hat S^+}{\partial \hat a} \hat a = \hat S^+ \, .
\label{splboson}
\end{equation}
Then, expanding $\hat S^+$ as
\begin{equation}
\hat S^+ = \sum^{\infty}_{m,n=0} S^{+}_{m,n} \hat a^{+m} \hat a^n \, ,
\label{splansatz}
\end{equation}
and inserting into Eq.~\eqref{splboson} directly gives $m = n + 1$. Therefore,
\begin{equation}
\hat S^+ = \hat a^+ \sum^{\infty}_{n=0} S^{+}_{n+1,n} \hat a^{+n} \hat a^n \equiv \hat a^+ \Phi^{+} \, .
\label{splresult}
\end{equation}
Besides the normal-ordered form provided by Eq.~\eqref{splresult}, $\Phi^+$ is unspecified. However, by virtue of Eq.~\eqref{maxsplboson},
\begin{equation}
\Phi^+ \vert 2S \rangle = 0 \, ,
\label{phiplconstr}
\end{equation}
must be fulfilled. Unfortunately, that crucial point is often overlooked.

The equation for $\hat S^-(\hat a^+, \hat a)$ can be derived in the similar manner from Eqs.~\eqref{spincommminz} and 
\eqref{szboson}:
\begin{equation}
\frac{\partial \hat S^-}{\partial \hat a} \hat a - \hat a^+ \frac{\partial \hat S^-}{\partial \hat a^+} = \hat S^- \, .
\label{sminboson}
\end{equation}
Inserting
\begin{equation}
\hat S^- = \sum^{\infty}_{m,n=0} S^{-}_{m,n} \hat a^{+m} \hat a^n \, ,
\label{sminansatz}
\end{equation}
in Eq.~\eqref{sminboson} leads to $n = m + 1$. As a result, $\hat S^-$ reads
\begin{equation}
\hat S^- = \sum^{\infty}_{m = 0} S^{-}_{m,m+1} \hat a^{+m} \hat a^m \hat a \equiv \Phi^- \hat a\, .
\label{sminresult}
\end{equation}
By definition $\hat a \vert 0 \rangle = 0$, and so, Eq.~\eqref{minsminboson} is satisfied. Thus, at this stage of calculations, $\Phi^-$ is the unspecified function given by Eq.~\eqref{sminresult}.

From Eqs.~\eqref{splresult} and \eqref{sminresult} it follows that 
\begin{equation}
\hat S^+ \hat S^- = \hat a^+ \Phi^+ \Phi^- \hat a \, .
\label{splphiplminsmin}
\end{equation} 
This expression suggests that it is more convenient to proceed further on the basis of Eq.~\eqref{ssqrplmin} instead of Eq.~\eqref{spincommplmin}. Indeed, Eqs.~\eqref{ssqrplmin}, \eqref{ssqrresult} and \eqref{splphiplminsmin} give:
\begin{equation}
\hat a^+ \Phi^+ \Phi^- \hat a = 2 S \, \hat a^+ \hat a - \hat a^{+2} \hat a^{2} \, .
\label{phiplmineq} 
\end{equation}
Then, it is easy to see that
\begin{equation}
\Phi^+ \Phi^- = 2 S - \hat a^+ \hat a \, .
\label{phiplminresult} 
\end{equation}
Thus, $\Phi^+$ and $\Phi^-$ can be determined by factorization of the right-hand side of Eq.~\eqref{phiplminresult}. 
In doing so Eq.~\eqref{phiplconstr} must be satisfied.

There exist multiple solutions to that problem. For example, the choice
\begin{equation}
\Phi^+ = \Phi^- = \sqrt{2 S - \hat a^+ \hat a} \, ,
\label{hprepr} 
\end{equation}
leads to the Holstein-Primakoff transformation \cite{holstein1940}. Although Eq.~\eqref{phiplconstr} is fulfilled, it should be noted that the normal-ordered expansion of $ \sqrt{2 S - \hat a^+ \hat a} $ reads \cite{louisell1964}:
\begin{equation}
\sqrt{2 S - \hat a^+ \hat a} = \sum_{m=0}^{\infty}\sum_{n=0}^m \frac{(-1)^n\sqrt{2S-(m-n)}}{(m-n)!n!} \hat a^{+m} \hat a^m.
\label{rootexp}
\end{equation}
In order for Eq.~\eqref{rootexp} to be useful it is necessary to truncate this expansion at some order. However, that is not consistent with Eq.~\eqref{phiplconstr}, and Eq.~\eqref{ssqrresult} is no longer satisfied in this case.

The conjugated Dyson-Maleev representation \cite{dembinski1964} is reproduced by
\begin{align}
& \Phi^+ = \sqrt{2S} \left( 1 - \frac{\hat a^+ \hat a}{2S} \right) \, ,  \nonumber \\
& \Phi^- = \sqrt{2S} \, .   \label{conjmaleev}
\end{align}
Note that, the expressions originally suggested by Maleev in Ref.~\cite{maleev1958} do not obey Eq.~\eqref{phiplconstr}, and therefore, should be rejected.

The boson representation of spin operators permits to establish deep connection with the symmetry considerations \cite{miller1972}.
In order to show that, let us choose $\Phi^-=-1$. Then we have:
\begin{align}
& \hat S^+ =  -\hat a^+(2S - \hat a^+\hat a ) \, , \nonumber \\
& \hat S^- = - \hat a \, , \nonumber \\
& \hat S^z = -S + \hat a^+ \hat a \, .\label{sugroup}
\end{align}
Moreover, since the operators ${\mbox{d}}/{\mbox{d}z}$ and $z$ realize the boson commutation relation
\begin{equation}
[\frac{\mbox{d}}{\mbox{d}z} \ ,z] = 1 \, ,
\label{fockcomm}
\end{equation}
we can make the substitutions $\hat a^+ \rightarrow z \  \mbox{and} \ a \rightarrow {\mbox{d}}/{\mbox{d}z}\ $ in
Eq.~\eqref{sugroup} to obtain
\begin{align}
& \hat S^+ =  -2Sz + z^2\frac{\mbox{d}}{\mbox{d}z}\, , \nonumber \\
& \hat S^- = - \frac{\mbox{d}}{\mbox{d}z}\, , \nonumber \\
& \hat S^z = -S +z \frac{\mbox{d}}{\mbox{d}z} \, .\label{sualgebra}
\end{align}
In the vector space of polynomials of degree not greater than $2S$
$$
\frac{z^{S + M}}{\sqrt{(S + M)!\,(S - M)!}} \, ,
$$
set of equations \eqref{sualgebra} represent Hermitian operators with respect to the following scalar product \cite{takhtajan2008}
\begin{equation}
(\varphi,g) = \frac{(2S+1)!}{\pi} \int_C \frac{\varphi(z)\overline{g(z)}}{(1+|z|^2)^{2S+2}}\mbox{d}^2z \, .
\label{innerpr}
\end{equation}
So, we see that Eq.~\eqref{sualgebra} give a well known realization of the representation of the Lie algebra $su(2)$ \cite{takhtajan2008, miller1972}.

Finally, the corresponding boson representation of spin operators for the spin ground state $ \vert +S, S \rangle$ can be derived in the similar manner and is given by: 
\begin{align}
& \hat S^+ = - \hat a \, , \nonumber \\
& \hat S^- =  -\hat a^+(2S - \hat a^+\hat a ) \, , \nonumber \\
& \hat S^z = S - \hat a^+ \hat a \, .\label{sugroupspl}
\end{align}
It should be noted that both set of equations \eqref{sugroup} and \eqref{sugroupspl} are needed in the quantum theory of antifferomagnetism, for example.

\textit{Acknowledgments.} This work is supported by the Georgian National Science Foundation and Centre National de la Recherche Scientifique (Grant No.~04/01).

\end{document}